# Optimization of the X-ray incidence angle in photoelectron spectrometers

Vladimir N. Strocov

*Swiss Light Source, Paul Scherrer Institute, CH-5232 Villigen-PSI, Switzerland*

Interplay of the angle dependent X-ray reflectivity and absorption with the photoelectron attenuation length in the photoelectron emission process determine the optimal X-ray incidence angle $\alpha_{opt}$ which maximizes the photoelectron signal. Calculations in a wide VUV through hard-X-ray energy range show that $\alpha_{opt}$ goes with energy progressively more grazing from a few tens of degrees at 50 eV to about one degree at 3.5 keV accompanied by the intensity gain increasing up to a few tens of times as long as the X-ray footprint on the sample stays within the analyzer field of view. This trend is fairly material independent. The obtained results bear immediate implications for design of the (synchrotron based) photoelectron spectrometers.

The X-ray photoelectron spectroscopy (XPS) experiment is in general characterized by a disparity of some two orders of magnitude between the relatively large X-ray attenuation depth and relatively small photoelectron escape depth. Most of the photoelectrons are excited therefore at the depth much larger they can elastically escape from, and on their way to vacuum dissipate in a series of inelastic scattering events to form only the cascade secondary electrons background carrying not much spectroscopic information. Obviously, the way to win the elastic signal will be to deposit more energy of the X-rays closer to the surface, which can be achieved by going to grazing incidence angles.

The process of X-ray excited production of photoelectrons has been casted into an exact numerical framework in the seminal work of Henke (Henke, 1972) already in 1972. He has shown that a significant increase of the elastic (no-loss) photoelectron yield can be achieved with grazing X-ray incidence angles approaching the total external reflection (critical) angle $\alpha_c$. These results have received further theoretical developments (Fadley, 1974) including generalization to multilayer structures (Chester & Jach, 1993; Fadley *et al.*, 2003) as well as extensive experimental verification (Hayashi *et al.*, 1996; Kawai *et al.*, 1995).

Here, we analyze the interplay between the X-ray reflectivity, X-ray absorption and photoelectron escape processes with a particular attention to the effects of the photon spot size with respect to the analyzer field of view (FOV). The optimal X-ray incidence angle to achieve the maximal XPS intensity gain is determined in a wide energy range from VUV- to hard-X-rays. These results bear immediate implications for optimization of the experimental geometry of (angle-resolving) XPS spectrometers.

*Formalism.* - We will first recap the basic formalism to describe the X-ray excited photoelectron current $I_{PE}(\alpha, \vartheta)$ as a function of the X-ray grazing incidence angle $\alpha$ and photoelectron emission angle $\vartheta$ relative to the surface normal, see Fig. 1 (*left*). According to Beer-Lambert law, the electromagnetic field intensity $S(x)$ in the media exponentially decreases with depth $x$ as $S(x) = A(\alpha)e^{-\frac{x}{d(\alpha)}}$, where $d(\alpha)$ is the electromagnetic field penetration depth (perpendicular to the surface) and $A(\alpha)$ is a normalization coefficient. The latter is defined by the condition that $\int_0^\infty S(x)dx$, expressing the total absorption in the media, obeys the complementarity principle and is thus proportional to 1-$R$, where $R$ is the X-ray reflection coefficient. Performing this integral and

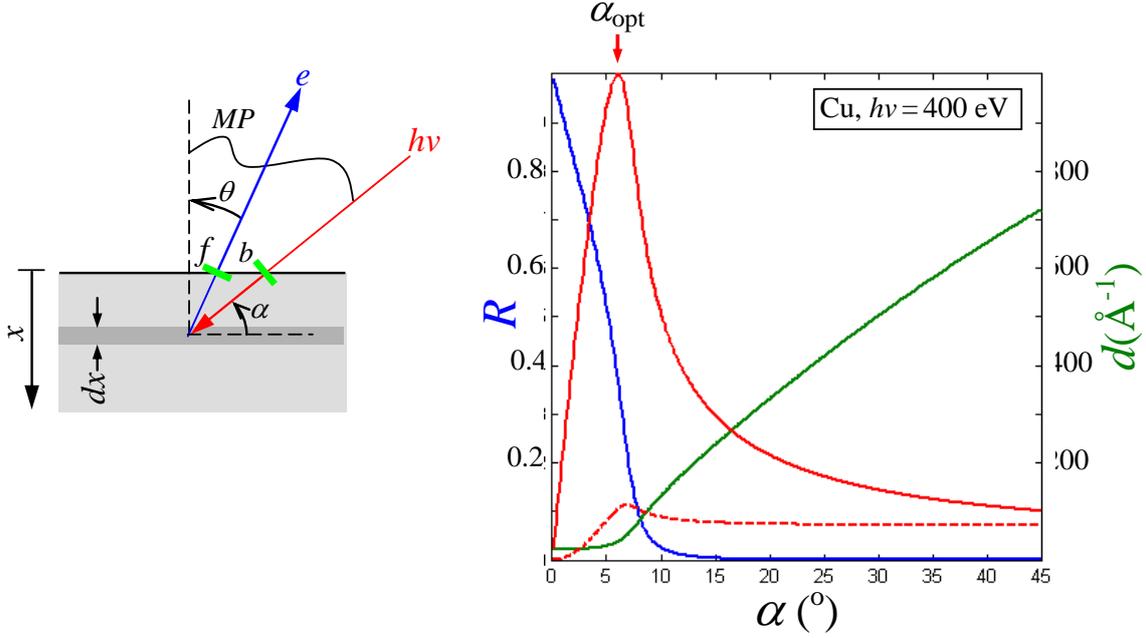

Fig. 1. (*left*) Sketch of the photoelectron emission process; (*right*) Angle dependences of the X-ray reflectivity $R$, absorption depth $d$ and the corresponding normal-emission $I_{PE}$ for Cu at $h\nu = 400$ eV in the FA (*solid line*) and FOVL (*dashed*) regimes. The decrease of $d$ combined with increase of $R$ towards more grazing $\alpha$ forms the peak of $I_{PE}$ prominent in the FA regime and only moderate in the FOVL one.

equating it to 1-$R$, we immediately obtain $A(\alpha) \propto \dfrac{1-R(\alpha)}{d(\alpha)}$ and

$$S(x) \propto \frac{1-R(\alpha)}{d(\alpha)} e^{-\frac{x}{d(\alpha)}} \tag{1}$$

Then, neglecting the photoexcitation matrix elements and the photoelectron refraction important only at low energies, the photoelectron intensity $dI_{PE}(\alpha,\vartheta)$ originating from a layer with a thickness $dx$ placed in depth $x$ is proportional to the power absorbed in the layer $S(x)dx$ multiplied by the photoelectron transmission through the overlayer $e^{-\frac{x}{\lambda \cos \vartheta}}$, where $\lambda$ is the photoelectron attenuation length (Powell *et al.*, 1999). Note that in contrast to $d(\alpha)$ taken perpendicular to the surface and having the meaning of depth, $\lambda$ is taken along the photoelectron path and has the meaning of length. Integration of $dI(\alpha,\vartheta)$ over the depth yields

$$I_{PE}(\alpha,\vartheta) \propto \frac{1-R(\alpha)}{d(\alpha)} \int_0^\infty e^{-\frac{x}{d(\alpha)}} e^{-\frac{x}{\lambda \cos \vartheta}} dx \tag{2}$$

which evaluates to

$$I_{PE}(\alpha,\vartheta) \propto (1-R(\alpha)) \frac{\lambda \cos \vartheta}{d(\alpha) + \lambda \cos \vartheta} \tag{3}$$

*View factor.* – The above formalism implied that the analyzer collected all photoelectrons emerging at the sample. Now, the above expression (3) should be multiplied by a geometrical view factor $V(\alpha,\vartheta)$ which is defined by the relation of the incident beam cross-section $b$ and the analyzer FOV $f$, see Fig. 1 (*left*), in their projections to the sample, $b/\sin \alpha$ and $f/\cos \vartheta$, respectively. Obviously, with less grazing $\alpha$ and more grazing $\vartheta$, when the inequality $b/\sin \alpha < f/\cos \vartheta$ takes place, we have full photoelectron acceptance and $V(\alpha,\vartheta)$ is identically equal to 1. In the opposite case, the acceptance is FOV-limited and $V(\alpha,\vartheta)$ is equal to the ratio of the FOV and beam projections:

$$V(\alpha,\vartheta) = \begin{cases} 1 & (full\text{-}acceptance\ regime) \\ \dfrac{f\sin\alpha}{b\cos\vartheta} & (FOV\text{-}limited\ regime) \end{cases} \quad (4)$$

The full-acceptance (FA) regime implies that all photoelectrons emerging throughout the X-ray footprint on the sample are intercepted the analyzer FOV. It is typical of the nowadays synchrotron sources delivering a beam focused to some 10 µm and below. The FOV-limited (FOVL) regime (often referred to as overfilled analyser slit) implies the loss of the photoelectrons outside the analyzer FOV. It is typical of the laboratory X-ray or older synchrotron sources with their spot being of the order of 1 mm. Our formalism for the FOVL regime is equivalent to that of Henke (Henke, 1972) who back in 1972 implied exactly this situation. An illustrative comparison of the FA and FOVL regimes can be found, for example, at http://goliath.emt.inrs.ca/surfsci/arxps/introcss.html.

*Numerical examples and analysis.* - We will now use the above formalism in practical calculations. We restrict ourselves to the normal emission $\vartheta = 0$. In this case the formulas (3-4) reduce to

$$I_{PE}(\alpha) \propto V(\alpha)(1-R(\alpha))\frac{\lambda}{d(\alpha)+\lambda} \quad (5)$$

with $V(\alpha)=1$ for the FA regime and $V(\alpha)=\dfrac{f}{b}\sin\alpha$ for the FOVL one. The calculations were performed for the paradigm metal Cu at $h\nu = 400$ eV. The numerical values of $R$ and $d$ were taken from the X-ray database readily availably on the Web (Henke *et al.*, 1993). The photoelectron energy was taken equal to $h\nu$ as relevant for the valence band XPS. The corresponding $\lambda$ was taken as the inelastic mean free path[1] and calculated according to the TPP-2M formula (Powell *et al*, 1999) using the NIST Standard Reference Database implemented in the program IMFPWIN (NIST, 2011). The FOVL calculations assumed $f = b$ and therefore $V(\alpha) = \sin\alpha$.

Fig.1 (*right*) shows the calculated $R(\alpha)$ and $d(\alpha)$ curves, as well as the $I_{PE}(\alpha)$ ones different for the FA and FOVL regimes. We will now discuss the general trends seen in this figure: (1) The region of less grazing $\alpha$ away from $\alpha_c$. Simple geometrical considerations give here $d(\alpha) \propto \sin\alpha$. Furthermore, $d(\alpha) >> \lambda$ and $R(\alpha) \sim 0$. This simplifies the formula (5) to $I_{PE}(\alpha) \propto V(\alpha)/\sin\alpha$. For the FA regime, $V(\alpha)=1$ and the remaining $1/\sin\alpha$ dependence reflects the gradual increase when going to more grazing angles of the power absorbed in the surface region where the photoelectrons are coming from. For the FOVL regime, the $V(\alpha) \propto \sin\alpha$ factor reflecting the analyzer FOV overfilling compensates this trend to constant $I_{PE}(\alpha)$, in agreement with the Henke's results (Henke, 1972). Therefore, the intensity gain with more grazing $\alpha$ can be achieved in this region only under the FA experimental conditions; (2) The region near $\alpha_c$. Here $d(\alpha)$ in (5) sharply reduces when going to more grazing angles to blow up $I_{PE}$. The counter-trend is the increase of $R$ to reduce the total absorption and thus $I_{PE}$. These opposite trends form the $I_{PE}$ peak identifying the optimal incidence angle $\alpha_{opt}$. In the FA regime, the intensity gain is really dramatic. In the FOVL regime, the $V(\alpha) \propto \sin\alpha$ factor much moderates the gain to a factor about 1.5, in agreement with the previous theoretical and experimental results (Henke, 1972; Kawai *et al.*, 1995)

---

[1] We note that the electron attenuation length in the crystalline media is in general limited only by the inelastic scattering. The elastic scattering appearing in the Boltzmann transport equation is superseded by the band theory where the Bloch electrons are formed by self-consistent multiple elastic scattering on the periodic potential and thereby propagate freely through the crystal, with the elastic attenuation being relevant only for evanescent Bloch waves in the band gaps (see, for example, (Capart, 1969)) important only at low energies (Barrett *et al.*, 2005).

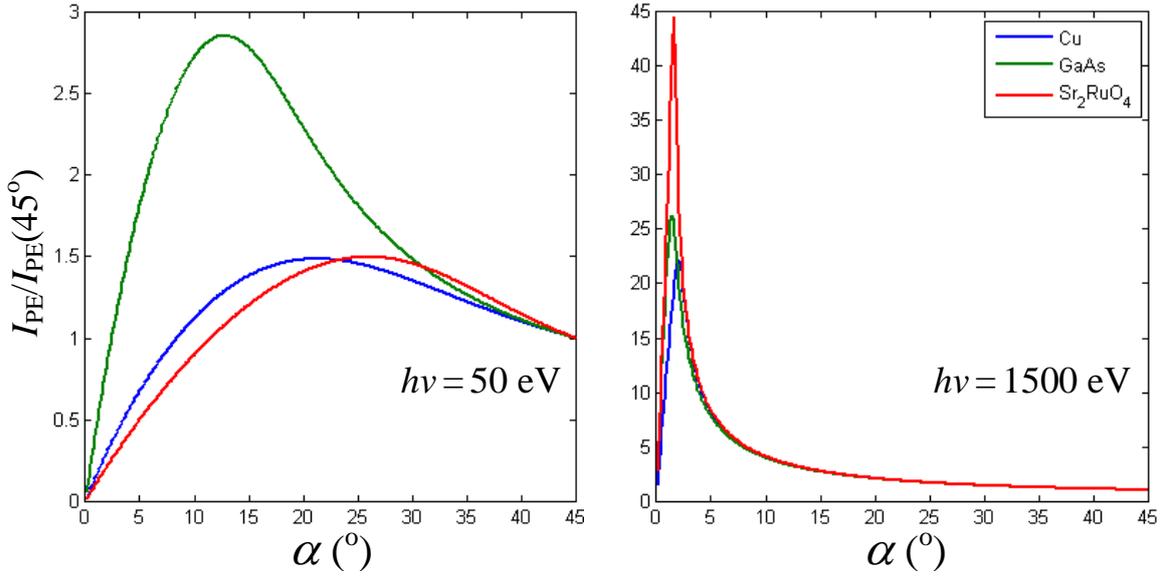

Fig. 2. Dependences of the normal-emission $I_{PE}$ on $\alpha$ (normalized to $I_{PE}$ at 45°) for three paradigm materials and two photon energies (FA regime). With increase of $hv$ the $I_{PE}$ peak moves to more grazing $\alpha$, dramatically scaling up in amplitude and sharpening.

and slightly shifts $\alpha_{opt}$ to less grazing angles. In the following, we will concentrate on the FA regime as more effective and relevant for the modern synchrotron instrumentation.

To assess the universality of the above picture, the calculations were extended to another two paradigm materials, the semiconductor GaAs and strongly correlated material $Sr_2RuO_4$, and to two very different $hv$ values, 50 eV and 1500 eV. The results are shown in Fig. 2. As we have seen above, the $R(\alpha)$ and $d(\alpha)$ angular dependences combined with $\lambda$ form the pronounced $I_{PE}$ peak at $\alpha_{opt}$ near $\alpha_c$. In the low-energy case, the peak appears at less grazing $\alpha$, it is broad and less pronounced compared to $I_{PE}$ at 45°. With increase of $hv$, the peak goes more grazing, dramatically scales up in amplitude and sharpens. These general trends appear fairly material independent.

*Optimal X-ray incidence angle.* – The above calculations were extended to determine the $\alpha_{opt}$ angle maximizing $I_{PE}$, see Fig. 1 (*right*), in a wide energy range from VUV photons of $hv = 50$ eV to hard X-rays of 3.5 keV.

Fig. 3 (*right*) shows the calculated energy dependence of $\alpha_{opt}$. It drops from less grazing values of the order of 20° at the 50 eV end to very grazing values about 1° at the 3.5 keV end. The physics of this behavior becomes clear from Fig. 3 (*left*) which presents the $hv$ dependence of $d$ taken at $\alpha$ = 45° far away from $\alpha_c$ (we note its sharp drops at the transition metal 2p absorption edges) compared with the energy dependence of $\lambda$. The increase of $d$ through the shown energy range is about an order of magnitude stronger than that of $\lambda$. The need to concentrate the absorbed X-ray power in a better balance with $\lambda$ forces $\alpha_{opt}$ to go more grazing with $hv$. Working in the same direction is also the evolution of $R(\alpha)$ whose onset shifts with $hv$ towards more grazing angles. Returning to Fig. 3 (*right*), we also note a dramatic increase with $hv$ of the intensity gain achieved at $\alpha_{opt}$.

Fig. 3 (*left*) also shows the $hv$ dependence of $d$ at $\alpha_{opt}$. Due to $\alpha_{opt}$ going progressively more grazing, $d(\alpha_{opt})$ flattens compared to $d$ at 45°. It is interesting to note that, somewhat counterintuitively, the $I_{PE}$ maximum at $\alpha_{opt}$ in general does not balance $d$ and $\lambda$. This manifests the modulating effect of $R(\alpha)$ which reduces the absorbed X-ray power towards more grazing $\alpha$. The balance between $d$ and $\lambda$ improves however at the high-energy end. In this case $d$ starts to affect the probing depth of the XPS experiment on equal footing with $\lambda$.

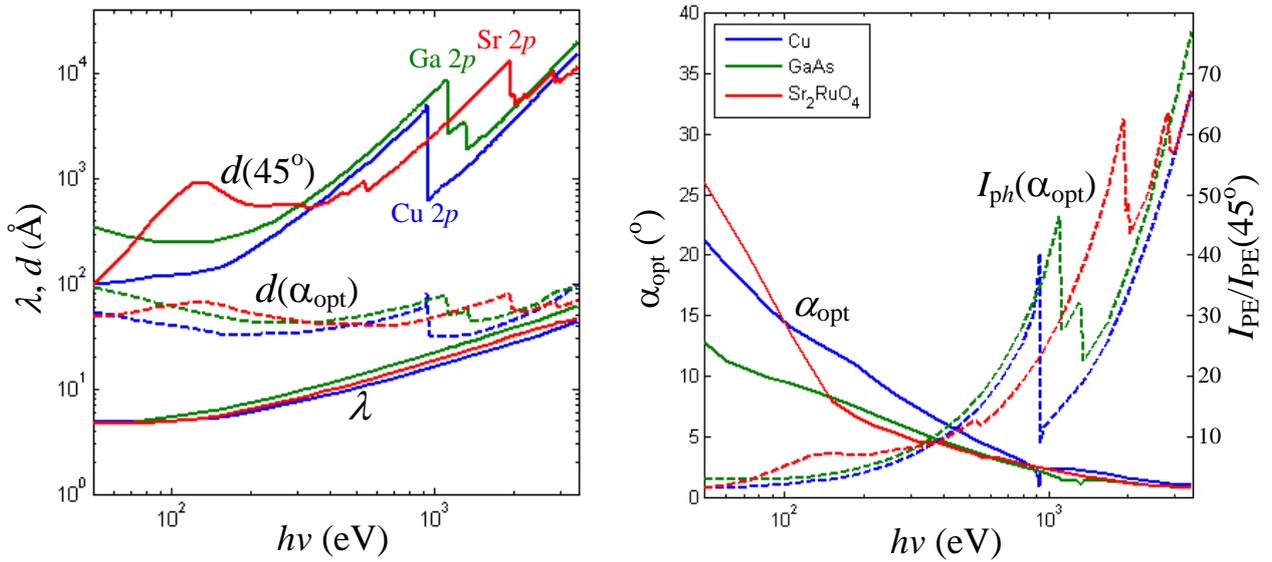

Fig. 3. (*left*) Energy dependences of $d$ (at $\alpha = 45°$ and the energy-dependent $\alpha_{opt}$ optimal angle) and $\lambda$; (*right*) Energy dependence of $\alpha_{opt}$ and the corresponding intensity gain compared to $\alpha = 45°$ (FA regime). Because $d$ increases with energy faster than $\lambda$, $\alpha_{opt}$ goes more grazing.

*Practical considerations.* – The above gigantic gain in $I_{PE}$ of a few tens of times looks like a miracle. However, it requires extremely grazing angles, and can only be fully realized as long as the analyser intercepts within its FOV the whole light spot blowing up proportionally to $(\sin\alpha)^{-1}$. Below we give some simple considerations to maintain or at least approach this FA regime at the grazing angles:

(1) Experimental geometry of the synchrotron based XPS facilities should take into account the elliptical cross-section of the incident beam with its relatively small vertical size $b_V$ and large horizontal size $b_H$. This means that the sample should be taken to the grazing incidence by rotation around the horizontal axis to blow up the smaller $b_V$ rather than the larger $b_H$. In this case the measurement plane (MP) formed by the incident beam and analyzer lens axis, see Fig. 1 (*left*), is vertical. This geometry has been implemented, for example, at the highly efficient soft-X-ray ARPES facility at the ADRESS beamline (Strocov *et al.*, 2010) of Swiss Light Source;

(2) The analyzer FOV is determined by the operation mode of the analyzer lens (Mårtensson *et al.*, 1994; Wannberg, 2009) and on the opening and orientation of the analyzer slit. For the magnification (essentially imaging) modes, the FOV is just the slit width $s$ divided by the lens magnification $M$: $f = s/M$. Obviously, the operation at grazing angles will benefit from low magnifications. Furthermore, the analyzer slit should be oriented in the MP. In this case the FOV will be determined by the relatively large slit length of the order of 20 mm. With the modern synchrotron sources and vertical MP, the FA regime is in this case practically unlimited in grazing angles. The transmission lens modes normally deliver even larger FOV compared to the magnification ones, but its determination does not obey to the simple imaging considerations because the photoelectrons collected at the same point at the slit can originate from different points on the sample. The angle-resolving XPS (ARPES) measurements are the most restrictive on the spot size, because the best angular resolution is ensured within a FOV of the order of 100 μm;

(3) Owing to a heavy weight of the high-resolution XPS analyzers, they are normally mounted at one of the flanges of the experimental vacuum chamber. The normal-emission $\alpha$ stays therefore fixed. It is reasonable to optimize it near the low-energy end of the required $h\nu$ range because for all other energies this $\alpha$ will sit on the less steep right side of the $R(\alpha)$ peaks, see Fig. 2. With the soft X-ray energy range starting around 300 eV, this yields $\alpha_{opt} \sim 8°$, see Fig. 3. For the ARPES measurements with $f \sim 100$ μm, the FA regime will then be limited by $b_V = f/\sin\alpha_{opt}$ of about 14 μm. This is hardly a problem for the nowadays synchrotron instrumentation. With the hard-X-ray energy range starting around 2.5 keV, we arrive at $\alpha_{opt} \sim 1.3^o$. For the ARPES measurements in this region (see, for example, Gray *et al*., 2011) the FA regime will be limited by $b_V \sim 2.3$ μm which requires already aggressive focusing of the incident beam.

We also note that an additional advantage of grazing $\alpha$ is a reduction of the inelastic secondary electron background, because the concomitant decrease of *d* reduces the secondary electron background originated from photoelectrons excited in the sample depth beyond $\lambda$ and inelastically scattered on their way to the surface (see, for example, Kawai *et al*., 1995).

*Conclusion.* – We have analysed the interplay between the X-ray reflectivity *R* and absorption depth *d* depending on the X-ray incidence angle $\alpha$ and the photoelectron attenuation length $\lambda$ in the photoelectron emission process. With increase of energy from the VUV- to hard-X-rays, the optimal X-ray incidence angle $\alpha_{opt}$ delivering maximal XPS signal goes progressively more grazing from a few tens of degrees to about $1^o$. This is accompanied by increase of the intensity gain at $\alpha_{opt}$ from insignificant to a factor of a few tens as long as the experiment stays in the FA regime with the whole X-ray footprint on the sample intercepted within the analyser FOV. The practical utilization of this gain by the synchrotron based XPS spectrometers in general requires vertical measurement plane with in-plane orientation of the analyzer slit and, in particular for ARPES experiments towards the hard-X-ray energies, focusing of the incident X-ray beam down to a few μm.

The author thanks G. Palsson, C. Cancellieri and C.S. Fadley for advice and valuable discussions.